\documentclass[twoside,a4paper]{article}
\pdfoutput=1

\usepackage[capposition=top]{floatrow}

\usepackage[sc]{mathpazo} % Use the Palatino font
\usepackage[T1]{fontenc} % Use 8-bit encoding that has 256 glyphs
\usepackage{microtype} % Slightly tweak font spacing for aesthetics

\usepackage[hmarginratio=1:1,top=32mm,columnsep=20pt]{geometry} % Document margins
\usepackage{multicol} % Used for the two-column layout of the document
\usepackage[hang, small,labelfont=bf,up,textfont=it,up]{caption} % Custom captions under/above floats in tables or figures
\usepackage{booktabs} % Horizontal rules in tables
\usepackage{float} % Required for tables and figures in the multi-column environment - they need to be placed in specific locations with the [H] (e.g. \begin{table}[H])
\usepackage[round]{natbib}         % bibliography style
\usepackage[colorlinks=true, citecolor=blue, urlcolor=blue]{hyperref}    % better urls in bibliography

\usepackage{lettrine} % The lettrine is the first enlarged letter at the beginning of the text
\usepackage{paralist} % Used for the compactitem environment which makes bullet points with less space between them

\usepackage{abstract} % Allows abstract customization
\usepackage{mathtools}
\usepackage{xfrac}
\usepackage{graphicx}
\usepackage{wrapfig}
\usepackage{lscape}
\usepackage[figuresright]{rotating}
\usepackage{epstopdf}
\usepackage{endnotes,etoolbox}
\let\footnote=\endnote

\patchcmd{\theendnotes}
  {\makeatletter}
  {\makeatletter}
  {}{}

\usepackage{titlesec} % Allows customization of titles
\titleformat{\section}[block]{\large\scshape\center}{\thesection.}{1em}{} % Change the look of the section titles
\titleformat{\subsection}[block]{\large}{\thesubsection.}{1em}{} % Change the look of the section titles

\usepackage{fancyhdr} % Headers and footers
\title{\vspace{-15mm}\fontsize{24pt}{10pt}\selectfont\textbf{The Spatial Structure of Transnational Human Activity\thanks{I am grateful to the participants of the BIGSSS Colloquium at Jacobs University Bremen for invaluable comments on an earlier version of this paper, to the World Tourism Organization for generously granting me access to their data on global tourism flows and to Lea Kliem for assistance in compiling the online friendship matrix used in this study.}}} % Article title

\author{
\large
\textsc{Emanuel Deutschmann}\\[2mm] % Your name
\normalsize Bremen International Graduate School of Social Sciences \\ \normalsize School of Humanities and Social Sciences, Jacobs University Bremen \\ % Your institution
\normalsize \href{mailto:e.deutschmann@jacobs-university.de}{e.deutschmann@jacobs-university.de} % Your email address
\vspace{-5mm}
}

\date{
\vspace{+5mm}
\small
Draft version, \today}

\begin{document}

\maketitle % Insert title

\thispagestyle{empty} % All pages have headers and footers

%----------------------------------------------------------------------------------------
%	ABSTRACT
%----------------------------------------------------------------------------------------

\begin{abstract}

\noindent Recent studies have shown that the spatial structures of animal displacements and local-scale human motion follow L\'{e}vy flights. Whether transnational human activity (THA) also exhibits such a pattern has however not been thoroughly examined as yet. To fill this gap, this article examines the planet-scale spatial structure of THA (a) across eight types of mobility and communication and (b) in its development over time. Combining data from various sources, it is shown that the spatial structure of THA can indeed be approximated by L\'{e}vy flights with heavy tails that obey power laws. Scaling exponent and power-law fit differ by type of THA, being highest in refuge-seeking and tourism and lowest in student exchange. Variance in the availability of resources and opportunities for satisfying associated needs appears to explain these differences. Over time, the L\'{e}vy-flight pattern remains intact and remarkably stable, contradicting the popular idea that socio-technological trends lead to a ``death of distance.'' Longitudinal change occurs only in some types of THA and predominantly at short distances, indicating regional shifts rather than globalization. 
\end{abstract}


%----------------------------------------------------------------------------------------
%	ARTICLE CONTENTS
%----------------------------------------------------------------------------------------

%\begin{multicols}{2} % Two-column layout throughout the main article text

\section{Introduction}

%\lettrine[nindent=0em,lines=3]{A} 
A hungry shark searching for prey in the ocean will frequently move short distances, interrupted by random changes of direction that are only occasionally followed by moves over longer distances. Sorting the displacement lengths covered in this journey by size and frequency results in a distribution with a long tail that obeys a power law\footnote{A power law describes the relationship between two quantities, where one quantity varies at the power of the other. Mathematically, the power-law relation between two quantities $y$ and $x$ can be defined as $y = ax^{-\beta}$, where a is a prefactor and $\beta$ is the scaling exponent. Power laws are common in the social and natural world and have been shown to exist in phenomena as diverse as the gamma-ray intensity of solar flares \citep{clauset2009power}, the sizes of strike waves \citep{biggs2005strikes}, and  the distribution of accusations between detainees \citep{deutschmann2014between}.}. Such mobility patterns, so-called L\'{e}vy flights\footnote{A L\'{e}vy flight is a random walk (i.e. a succession of steps into random directions) in which the step-lengths feature a heavy-tailed probability distribution. For a precise mathematical definition, cf. \citet{shlesinger1986levy}. In empirical research, L\'{e}vy flights are diagnosed by ``showing that the power-law distribution holds'' \citep[p. 715]{buchanan2008mathematical}.}, have been shown to occur not only in the foraging movements of sharks and other marine predators like sea turtles and penguins \citep{sims2008scaling}, but also in the motion of smaller species like plankton \citep{bartumeus2003helical}. For humans, L\'{e}vy-like patterns have been found in local \citep{gonzalez2008understanding, song2010modelling, rhee2011levy}, and nation-wide mobility \citep{brockmann2006scaling}. Whether \textit{planet-scale} human activity also follows a L\'{e}vy flight has however not been thoroughly examined to date.

While this natural-scientific strand of research on L\'{e}vy flights has largely omitted the global sphere as yet, the social-scientific literature that deals with the global sphere has conversely not taken the findings of the L\'{e}vy-flight debate into account. On the contrary, it is dominated by the idea that technological and socioeconomic trends have led to a diminishing or even vanishing role of physical distance in structuring human activity. Symptomatic are shibboleths like ``demise of geography'' \citep{toffler1970future}, ``time-space compression'' \citep{harvey1989condition}, ``end of geography'' \citep{o1992global}, ``collapse of space'' \citep{kirsch1995incredible}, ``death of distance'' \citep{cairncross1997death}, and ``flat world'' \citep{friedman2007world}. Although a third stream of research has argued that such statements are exaggerated and that geographic space (still) structures social relations, a rigorous empirical analysis of the association between physical distance and planet-scale human activity is also missing in the social sciences. 

The persistence of this research gap and the stark contrast between the natural- and social-scientific positions is astonishing given that many problems humanity is facing today, from health epidemics like H5N1 and Ebola to climate change and terrorism, are heavily intertwined with transnational human mobility and communication. Driven by the hope that a better understanding of how people move around the world can contribute to solving such pressing social problems, a group of researchers has recently challenged the scientific community to get active and ``to collect large-scale human mobility traces'' \citep{hui2010planet}. The aim of this article is to respond to this call and to fill in some missing pieces, namely to add the global-scale analysis to the natural-scientific L\'{e}vy-flight debate, and the L\'{e}vy-flight analysis to the social-scientific globalization debate. In specific, we search for answers to the following research questions:
\begin{enumerate} \itemsep1pt \parskip0pt
\parsep0pt
  \item Does the spatial structure of transnational human activity follow a L\'{e}vy-flight?
  \item Does the spatial structure differ by type of activity, and if yes --- why?
  \item Has the spatial structure of transnational human activity changed over time?
\end{enumerate}
To do so, data on five types of mobility (asylum-seeking, migration, refuge-seeking, student mobility, tourism) and three types of communication (phone calls, online friendships, remittances) from a range of sources was standardized to cover the same set of 196 sending and receiving countries (i.e. 38,220 country dyads). In total, the structure of approximately 1.8 million valued ties across 47 network matrices at various time points between 1960 and 2010 are analyzed, involving, inter alia, 1.7 billion tourist trips and 240 billion international phone call minutes. While the online friendship dataset for instance is entirely new and thus original in itself, the major innovation of this study is to link the hitherto disconnected natural- and social-scientific debates by conducting the first encompassing comparative analysis of the spatial structure of transnational human activity (a) at a planet-wide scale, (b) across various types of mobility and communication, and (c) over time.

This article is organized as follows: in the next section, we review the existing literature and theories concerning the spatial structure of human (and non-human) activity. We then set out our own conceptual approach. Next, the data and methods are described. Finally, we present the results, focusing first on the current spatial structure, then move to examining longitudinal trends, and finally compare our results to theoretical scenarios and results from other studies. We conclude with a summary of our findings and a discussion of their implications. \\

\section{Theory and state of research}

The literature on the spatial structure of human activity can be described as consisting of three separate streams: (a) the L\'{e}vy-flight debate, (b) the geography-is-dead debate, and (c) the distance-decay debate. In the following, we briefly review these streams of research, highlighting their achievements and deficits. 

\subsection{The L\'{e}vy-flight debate}
The L\'{e}vy-flight debate is carried out mainly by natural and complexity scientists and began with research that showed that the motion patterns of many non-human species, from mammals like spider monkeys \citep{ramos2004levy} to marine predators like sharks, sea turtles and penguins \citep{sims2008scaling} and much smaller organisms like plankton \citep{bartumeus2003helical}, follow power laws. While some of the early alleged L\'{e}vy-flight diagnoses (e.g.\ for albatrosses) turned out not to hold upon closer scrutiny \citep{buchanan2008mathematical}, the debate remains vital and was recently expanded to human mobility. L\'{e}vy-like mobility patterns have now been found in studies that followed the GPS traces of volunteers on university campuses, theme parks, state fairs, and in cities \citep{rhee2011levy, noulas2012tale}, the movements of mobile phones \citep{gonzalez2008understanding, song2010modelling}, and the traces of banknotes in the United States \citep{brockmann2006scaling}. 

The theoretical argument for the L\'{e}vy-flight pattern is --- at least for animal motion --- the optimization of food search. It has been shown that in environments where food is scarce, a power-law distribution of flight lengths with a scaling exponent $\beta$=2 is the optimal search strategy \citep{viswanathan1999optimizing}. For human mobility, though, the existing papers contain little theoretical reflection as to \textit{why} the L\'{e}vy flight should apply (e.g. whether the goals humans pursue in their everyday life are readily equitable with foraging). Another critical point is that in contrast to the random walk that underlies the L\'{e}vy-flight model, humans do not actually move into random directions. It is more likely that the directions are determined by needs humans seek to fulfil \citep{noulas2012tale} or the information they have about their environment \citep{miller1972note}. Yet, while this discrepancy has repeatedly been recognized \citep{rhee2011levy, song2010modelling}, little has been done to overcome it. The biggest shortcoming of the L\'{e}vy-flight debate however are its current scale and data restrictions: US-dollar bills for instance can only be used in the United States and ``it remains unclear whether the observed properties are specific to the US or whether they represent universal features'' \citep[p. 33]{brockmann2008money}. Two recent studies \citep{cheng2011exploring, noulas2012tale} have shown that log-ins to location sharing services (LSS) like Foursquare follow a L\'{e}vy-flight pattern on a global scale. Yet, LSS require smartphone access and are used only by a small, well-off minority of the world population. Mainstream forms of planet-scale mobility and communication (including poverty-driven ones) are disregarded by the L\'{e}vy-flight debate as yet, as is potential longitudinal change. \\

\subsection{The geography-is-dead debate}
The idea of a decline and eventual vanishing of the role of space in structuring human interaction has a long history and takes many facets. Already in the mid-19th century, Marx and Engels argued that ``[i]n place of the old local and national seclusion and self-sufficiency, we have intercourse in every direction, universal interdependence of nations'' (\citeyear[p. 12]{marx1948manifesto}). Later, Marx coined the term ``annihilation of space by time'' to describe the tendency of capital (and in its wake all parts of society) to ``tear down every spatial barrier to intercourse'' (\citeyear[p. 539]{marx1973grundrisse}). In 1962, McLuhan argued that the ``new electronic interdependence recreates the world in the image of a global village'' (\citeyear[p. 31]{mcluhan1962gutenberg}). A decade later, Toffler coined the term ``demise of geography'' arguing that in contrast to the nomads of the past who were bound by place, ``the new nomads of today leave the physical structure behind'' (\citeyear[p. 91]{toffler1970future}). Similarly, O'Brien argued for an ``end of geography'' (\citeyear{o1992global}) in financial markets, while Cairncross maintained that ``[t]he death of distance as a determinant of the cost of communicating will probably be the single most important force shaping society in the first half of the next century'' (\citeyear[p. 1]{cairncross1997death}). Friedman stated that we recently entered a new era of globalization which ``is shrinking the world from a size small to a size tiny and flattening the playing field'' (\citeyear[p. 10]{friedman2007world}). The central characteristic of this new era, he argued, was ``the newfound power for individuals to collaborate and compete globally'' (Ibid). Numerous other terms, such as ``time-space compression'' \citep{harvey1989condition}, ``collapse of space'' \citep{kirsch1995incredible}, and ``shrinking world'' \citep{allen1995shrinking} have been proposed to describe this phenomenon.

The recurrent argument in this debate is that technological innovations, declining costs and massive growth of transport and communication infrastructure lead to a world in which physical distance plays a smaller or even no role at all anymore, first and foremost for flows of money and messages, but also for the movement of commodities and individuals. The geography-is-dead debate in its current and past forms has largely been driven by theoretical arguments with little empirical grounding. Yet its arguments have been highly influential in the social sciences and popular in the public sphere. It is completely detached from the L\'{e}vy-flight debate, but another stream of more empirics-oriented research, the distance-decay debate, has --- explicitly and implicitly --- referred to it critically. \\

\subsection{The distance-decay debate}
This third stream of research acknowledges the (continuing) salience of distance in structuring human activity, however usually without analyzing the precise shape of this relation as in the L\'{e}vy-flight debate. Most studies do not go beyond the notion that physical proximity (still) matters. A typical statement, which became known as Tobler's First Law of Geography, is: ``Everything is related to everything else, but near things are more related than distant things'' \citep[p. 236]{tobler1970computer}. Other central terms are the ``principle of least effort'' \citep{zipf1949human} and ``distance decay''. ``Distance decay'' is used in a wide range of fields from criminology, where it is known that offenders tend to commit crimes in proximity to their residency \citep{rengert1999distance}, to eco-geography, where studies found that biological similarity decreases with geographical distance \citep{soininen2007distance}. In military science, the continued importance of the ``loss of strength gradient'' has been emphasized \citep{webb2007continued}. Several studies on international trade and transport costs have explicitly criticized the idea of a ``death of distance'' as exaggerated \citep{kano2013exaggerated}, unfounded \citep{leamer1995international}, or even maintained the opposite: that instead of diminishing or disappearing, the effect of distance on trade rather \textit{increased} during the 20th century \citep{disdier2008puzzling}.\footnote{In contrast to the L\'{e}vy-flight debate's assumption of a non-linear power-law relation, these economic studies seem to assume a linear relation (e.g., when describing the influence of distance on trade in the form of a single ``distance coefficient'' or ``elasticity''), as such mean effect sizes would otherwise be rather meaningless.} 

With respect to human mobility, \citet{ravenstein1885laws} already studied the ``laws of migration'' using census data from England in the 19th century. Although his contemporaries thought that the term ``laws'' was inappropriate (Ibid, p. 233), he realized that the large majority of migrants only moved short distances while few migrants moved long distances. Later, \citet{stouffer1940intervening} presented graphically a spatial distribution of family movements in Cleveland that featured a heavy tail. Concerning \textit{transnational} human activity, the central structuring role of geographical distance has been recognized in a number of empirical studies. For transactions like diplomatic exchange and mail flows, geographic proximity was described as ``the most compelling force of attraction in the international system'' \citep[p. 889]{brams1966transaction}, ``a very pronounced influence and constraint on the pattern of West European communications and interactions'' \citep[p. 226]{clark1987european} and ``one of the major factors in global communication networks'' \citep[p. 181]{choi1995effect}. \citet{mckercher2008impact} used the concept of ``distance decay'' in relation to international tourism. Focusing on 41 countries, they show that 80 percent of all international travel occurs to countries within 1,000km distance. Pertaining to the supposedly placeless digital world, a recent study on the spatial structure of the internet found an effect of physical distance in the form of a power law with a cutoff \citep{tranos2013death}, while another study only reported that the structure of transnational Facebook friendships was ``apparently influenced by geography'' \citep[p. 13]{ugander2011anatomy}. \citet[p. 78]{takhteyev2012geography} showed a distribution of Twitter messages by geographical distance in which distance clearly mattered, however again without analyzing this relation further.

Two explanations have been put forward as to why human mobility is affected by geographic space \citep[cf.][]{miller1972note, noulas2012tale}. The first one, the \textit{gravity} hypothesis, is inspired by Newton's law of gravity and states that costs connected to distance itself are responsible for fewer long distance movements \citep[cf.][]{deutsch1961note, zhou2011intensification}. The second one, the \textit{intervening opportunities} hypothesis, argues that it is not the costs of distance itself that matter, but intervening opportunities that allow to fulfill one's needs already at close distances, making long-distance mobility unnecessary \citep[cf.][]{stouffer1940intervening, freymeyer1985spatial}. 

While in many ways more differentiated than the geography-is-dead debate, an encompassing analysis of the precise shape of the relation between distance and various types of human activity on a global scale is also missing in this stream of research. Moreover, to our knowledge no study has tracked change in the spatial structure of human mobility and communication over time. \\

\section{Conceptual approach}
Our aim is to analyze and compare the spatial structure of various types of transnational human activity (THA). We use THA as an umbrella term for transnational human mobility (THM), which denotes activities in which individuals cross nation-state borders physically, and transnational human communication (THC), which comprises communicative acts across nation-state borders that do not necessarily involve physical mobility. Starting from the L\'{e}vy-flight debate, we first test to which extent various types of THA feature a probability distribution whose tail follows a power-law function of the form 

\begin{equation}
P(r) \propto r^{-\beta},
\end{equation}

where $P(r)$ is the probability of a displacement length $r$ to occur and $\beta$ is the scaling exponent. A larger $\beta$ means that the curve is steeper, i.e. relatively more activity occurs at short distances, whereas a smaller $\beta$ denotes a flatter relation, i.e. relatively more activity takes place over longer distances. We expect $\beta$ and the fit of this power-law function (measured as $R^2$) to vary by type of THA. For THM, the logic of our argument, which is inspired by the distance-decay debate (and in specific the intervening opportunities hypothesis), is the following: if we accept the assumptions that

\begin{enumerate} \itemsep1pt \parskip0pt
\parsep0pt
  \item THM is associated with type-specific goals, 
  \item the availability of opportunities for goal-attainment varies by goal 
  \item the average amount of resources available to attain goals varies by type of THM, 
  \item humans aim at spending as little of their resources as necessary to attain their goals, 
\end{enumerate}

then the spatial structure of a specific type of THM $i$ can be expected to be determined by two factors: the availability of opportunities for attainment of the goals associated with $i$ and the resources available on average to the individuals engaging in $i$. The broader the availability of opportunities for goal-attainment, the higher is the likelihood for individuals to stop their movement at closer locations (and thus the higher $R^2$ and $\beta$), because their needs are already fulfilled to a satisfying extent and any further movement would only diminish the stack of resources without leading to additional benefits. Similarly, the higher the average amount of resources available to the group of people engaging in a particular type of THM, the less they will be physically bound by the costs of THM (and thus the lower $R^2$ and $\beta$), ceteris paribus. The real world is of course more complex, but still this simple model may help to obtain a first explanation for systematic differences between spatial structures of various THM types.

Our considerations allow to delineate specific expectations for the five types of THM under study (Table~\ref{table:tableone}). Major goals commonly associated with refuge- and asylum-seeking are mere survival, fulfillment of basic needs and security. These goals can usually be fulfilled in many places, oftentimes already in neighboring countries just outside a warzone. The resource stock available to refugees and asylum-seekers tends to be rather low. As a result, the spatial structure of refuge- and asylum-seeking is expected to feature a high power-law fit and scaling exponent. Tourists are often interested in a pleasant, entertaining environment (if on holidays) or in business opportunities (if on business trips). Both are widely available in many countries around the world (and easier to pursue in closer ones, e.g.\ due to cultural similarity), but tourists are likely to possess more resources on average than refugees. Therefore, $R^2$ and $\beta$ should still be high for tourism but slightly lower than for refuge-seeking. Migrants are often interested in improving their economic well-being, which is quite stratified globally and although moving short distances may already result in relative improvements, moving a bit further may in many cases still lead to additional benefits. Migrants are also likely to possess more resources on average than refugees, so that overall we expect a medium $R^2$ and $\beta$. International students tend to aim for excellent education and social distinction, which is best available only in a small number of institutions in a select number of countries, as the global university system is highly stratified \citep{barnett1995international}. In theoretical terms, this stratification leads to a lack of intervening opportunities: students cannot just go to neighboring countries but have to reach England or the US to attain their goals. Therefore --- and because international students will also be comparatively well-situated on average --- we expect low $R^2$- and $\beta$-values for student exchange.

\begin{table}[t]
%\def\~{\hphantom{0}}
%\begin{minipage}{168mm}
\caption{Theoretical expectations concerning THM}
\label{table:tableone}
\centering
\begin{tabular}{p{70pt}p{100pt}p{43pt}p{43pt}p{43pt}p{43pt}}
\toprule
  & Major goal(s) & Ubiquity & Resources & Power-law fit ($R^2$) & Scaling exponent ($\beta$) \\
\midrule
Refugees and asylum-seekers		& Survival, fulfilment of basic needs, security			& High 	&  Low		& High		& Large \\
Tourists 			& Entertainment, pleasant environment/ business opportunities	& High 	& Medium	& Medium-high	& Medium-high \\
 Migrants			& Economic well-being 			& Medium	&  Medium	& Medium	& Medium	\\
 Students		&  Education, distinction			& Low 	& High 		& Low  	& Small \\
 
\bottomrule
\end{tabular}
%\end{minipage}
\end{table}

For THC, our argument is inspired by the geography-is-dead debate. First of all, the spatial structure of THC will to some extent be a function of THM, as people will often communicate with friends, kin or business partners that have gone abroad. Yet, over and above, there may be variance inherent to the specific type of THC that depends on its technological standard and usage cost structure. This argument can easily be explained by comparing phone calls with online friendships. The two networks are similar in that both should to a certain extent resemble THM. Yet, they are different in that the monetary costs of international telephone communication are relatively high and increase with distance \citep[p. 6]{cairncross1997death}, while the material costs of online friendships are low and independent of distance: for someone from Switzerland, having an online friendship with an Austrian is as cheap as having one with an Australian, but calling Australia by phone is far more expensive than calling Austria. Accordingly, we expect the spatial structure of online friendships to be ``flatter'' and more detached from the power-law pattern (reflected in lower $\beta$ and $R^2$ values) than the spatial structure of phone calls. Remittances are special in that they are additionally influenced by economic power and we refrain from formulating a specific hypothesis about its spatial structure ex ante.

In the longitudinal part of the analysis, we build on the geography-is-dead debate's arguments in expecting the world to become ``flatter'' with globalization, and thus the type-specific $R^2$- and $\beta$-values to decline over time. In a final analytical step, we contrast the empirical THA distribution with three ideal geography-is-dead scenarios.  
 \\\\\

\section{Research design}

\subsection{Data}
This study is based on directed network matrices for eight types of THA at various time points between 1960 and 2010 as well as an undirected matrix for geographic distance, all of which were standardized\footnote{The set of 196 countries in the standardized matrices was obtained by excluding countries that were not contained in the original datasets of all types of THM under study. For some types of THC, the original set of countries was smaller and the matrices were artificially expanded. We decided to follow this inclusive approach as a more exclusive one would have resulted in a small, limited remainder of countries that would hardly allow speaking of a planet-scale analysis anymore. We set missing values to zero in all matrices, thereby following the example of \citet[p. 154]{reyes2013structure}, who showed the robustness of this approach for the UNWTO tourism dataset and finds that alternative procedures like multiple imputation lead to similar outcomes. In order to make the student and phone-call matrices from earlier years comparable with recent ones, historic states were replaced with their currently existing ``equivalents''. For example, Dahomey was equated with Benin, Upper Volta with Burkina Faso, and so on (full list of equations available upon request). This procedure does obviously not do justice to the complexity of historic developments, but what matters for the purposes of this article is a similar geographic location, not precise historic equivalence. The World Bank migration data, which also goes back to 1960, already comes in the form of currently existing states for all years in its original format.} to cover the same set of 196 sending and receiving countries (full list available upon request). Each of the 53 THA matrices thus contains information on $196 \times 196 - 196 =38,220$ country dyads. For the measurement of \textit{geographic distance} we draw on CEPII's GeoDist dataset \citep{mayer2011notes}. We use the weighted geodesic distance (\textit{distwces}), which provides the average distance between countries based on the spatial distribution of the population in the countries' 25 largest cities. Table~\ref{table:tabletwo} provides a summary of the eight types of THA under study. 

\begin{table}[t]
\caption{Types of THA used}
\label{table:tabletwo}
\centering
\begin{tabular}{p{20pt}p{85pt}p{38pt}p{38pt}p{45pt}p{110pt}}
\toprule
  & THA Type & Individual weight (\%) & Combined weight (\%) & Available years & Data source(s) \\
\midrule
THM		& Asylum seekers 		& $0.1$ 	&  		& 2010 		& \citet{unhcr2013population} 		\\
 		& Migrants 			& $16.9$ 	& 		& 1960-2010 	& World Bank, \citet{un2012migrant} 	\\
 		& Refugees 			& $0.8$ 	&  $60$	& 2000-2010 	& \citet{unhcr2013population}		\\
 		& Students 			& $0.2$ 	&  		& 1960-2010  	& PINA, \citet{unesco2013students}	\\
 		& Tourists 			& $82.0$ 	&  		& 1995-2010 	& \citet{unwto2014compendium}		\\
		& 				&  		&  		&  			&		 				\\
THC 		& Online friendships  	& $33.3$ 	& 		& n.d. 		& \citet{facebook2012mapping}		\\
		& Phone calls 		& $33.3$ 	&  $40$ 	& 1983-1995 	& PINA, ITU  				\\ 
		& Remittances 		& $33.3$ 	& 		& 2010 		& World Bank  				\\
\bottomrule
\end{tabular}
\end{table}

Of the eight types of THA, five involve physical mobility (THM): asylum-seeking, migration, refuge-seeking, student exchange, and tourism. Data on \textit{asylum-} and \textit{refuge-seeking} was obtained from UNHCR. According to the 1951 Refugee Convention (as broadened by a 1967 Protocol), a refugee is defined as a person who: 

\begin{quotation}
``owing to well-founded fear of being persecuted for reasons of race, religion, nationality, membership of a particular social group or political opinion, is outside the coun-try of his nationality and is unable or, owing to such fear, is unwilling to avail himself of the protection of that country; or who, not having a nationality and being outside the country of his former habitual residence as a result of such events, is unable or, owing to such fear, is unwilling to return to it.'' \citep{unhcr2014refugees}
\end{quotation}

An asylum-seeker in turn is ``someone who says he or she is a refugee, but whose claim has not yet been definitively evaluated'' \citep{unhcr2014asylum}. Data on \textit{migration} was extracted from the World Bank's Global Bilateral Migration Dataset for the years 1960 to 2000 \citep{ozden2011earth}, supplemented by United Nations data for the year 2010 \citep{un2012migrant}. The latter source defines migrants as ``foreign-born'' persons, or, where data on place of birth is unavailable, as ``foreign citizens'' \citep[p. 3]{un2012migrant}. Information on transnational \textit{student mobility} was obtained from Princeton's International Networks Archive (PINA)\footnote{Available at \url{http://www.princeton.edu/~ina/}, accessed 13/08/2013.} for the years 1960 to 1998 and from UNESCO for the years 2000 to 2010. UNESCO defines international students as ``[s]tudents who have crossed a national or territorial border for the purposes of education and are now enrolled outside their country of origin'' \citep[p. 264]{unesco2010global}. Data on \textit{tourism}, available from 1995 to 2010, was obtained from the World Tourism Organization (UNWTO), according to which ``[a] visitor (domestic, inbound or outbound) is classified as a tourist (or overnight visitor) if his/her trip includes an overnight stay'' \citep{unwto2008international}. Here, we are interested in ``arrivals of non-resident tourists at national borders, by country of residence''\footnote{For a few countries this category is unavailable. In order not to lose these countries, the category ``arrivals of non-resident \textit{visitors} at national borders'' was used in these instances. In cases where both these categories are missing, the category ``arrivals of tourists \textit{in all types of accommodation establishments}'' was used instead.}. Note that this definition does not premise any specific visiting purpose and may thus include business travel just as holiday trips.  

Three types of THA under study represent indirect communication (THC): online friendships, phone calls, and remittances. \textit{Online friendships} are based on Facebook data which was retrieved from an interactive graph that is available online \citep{facebook2012mapping} and converted into a network matrix. For each country $c$, this matrix contains the five countries with which $c$'s population has most Facebook friendships with, ranked from 5 (highest number of Facebook friendships) to 1 (fifth-highest number of Facebook friendships). Data on international \textit{phone calls} (measured in million minutes) from 1983 to 1995 originates from the International Telecommunication Union (ITU) and was retrieved from PINA \citep[cf.][]{louch1999phone}. Information on \textit{remittances} was obtained from the World Bank \citep[cf.][]{ratha2007south}. Remittances can be defined as ``current private transfers from migrant workers who are considered residents of the host country to recipients in the workers' country of origin'' \citep[p. xvi]{worldbank2011remittances}. We regard remittances as a type of THC because they are transfers between individuals that ``often involve related persons'' \citep[p. 75]{imf2005balance} and can thus be understood as expressions of support or solidarity, and ultimately as a form of communication. 

In addition to analyzing these eight activity types individually, we are also interested in getting an idea of what the spatial structure of THA looks like \textit{as a whole}. The multiplexity of human mobility and communication, i.e. the variety of ways in which people interact, needs to be addressed as concentrating on single activity types alone may lead to biases \citep{stopczynski2014measuring}. To do so, we link the activity types in three indices. First, a \textit{THM index}, in which the cell values of the 2010 matrices of the five types of mobility are added up. This simple procedure is reasonable because all mobility networks are based on the same unit of analysis (individuals moving between countries). As shown in Table 2, the weight in the THM index differs drastically by mobility type, with tourists and migrants making up for 82.0 and 16.9 percent, respectively, whereas asylum seekers, refugees\footnote{In some countries' census statistics, refugees are also counted as migrants, while in others they are not \citep{un2012migrant}. The UN tries to account for this by including refugees in ``most developing countries'' as migrants in the migration dataset (Ibd). It is therefore possible that refugees are sometimes counted twice. However their small relative weight (0.8 percent) shows that this issue has little practical consequences for the overall THM index.}, and students taken together account for only 1.1 percent of all THM. Second, a \textit{THC index} is created from the latest available matrix of the three forms of communication under study. This is less straightforward as the units differ between the types of THC (remittances are in US-Dollar, phone calls in minutes, etc.). We deal with this issue by normalizing the units and calculating the average value across the three types of THC, giving each of them the same weight. Third, we create a \textit{THA index} by adding the standardized values of THM and THC, giving a weight of .6 to the former and a weight of .4 to the latter. The purpose of these factors is to account for the fact that physical mobility requires more effort than indirect communication and should therefore receive more weight. The overall indices should be understood as only providing a tentative impression of THM, THC, and THA as a whole, because (a) we do not include all conceivable activity types, (b) the size of the weighting factors in the latter two indices is to a certain extent arbitrary, and (c) not all elements date from the same year (although our finding of long-term stability [see below] indicates that older data can readily be used as a proxy). Despite these shortcomings, we think that our indices constitute a significant first step to covering the multiplex nature of THA.
 \\

\begin{figure}[p]

  \caption{Probability density functions of THA and their power-law fit}
  \label{fig:one}
%\floatfoot{\emph{Note}: THA=transnational human activity, THM=transnational human mobility, THC=transnational human communication.}
  \centering
  \includegraphics[width=1\textwidth]{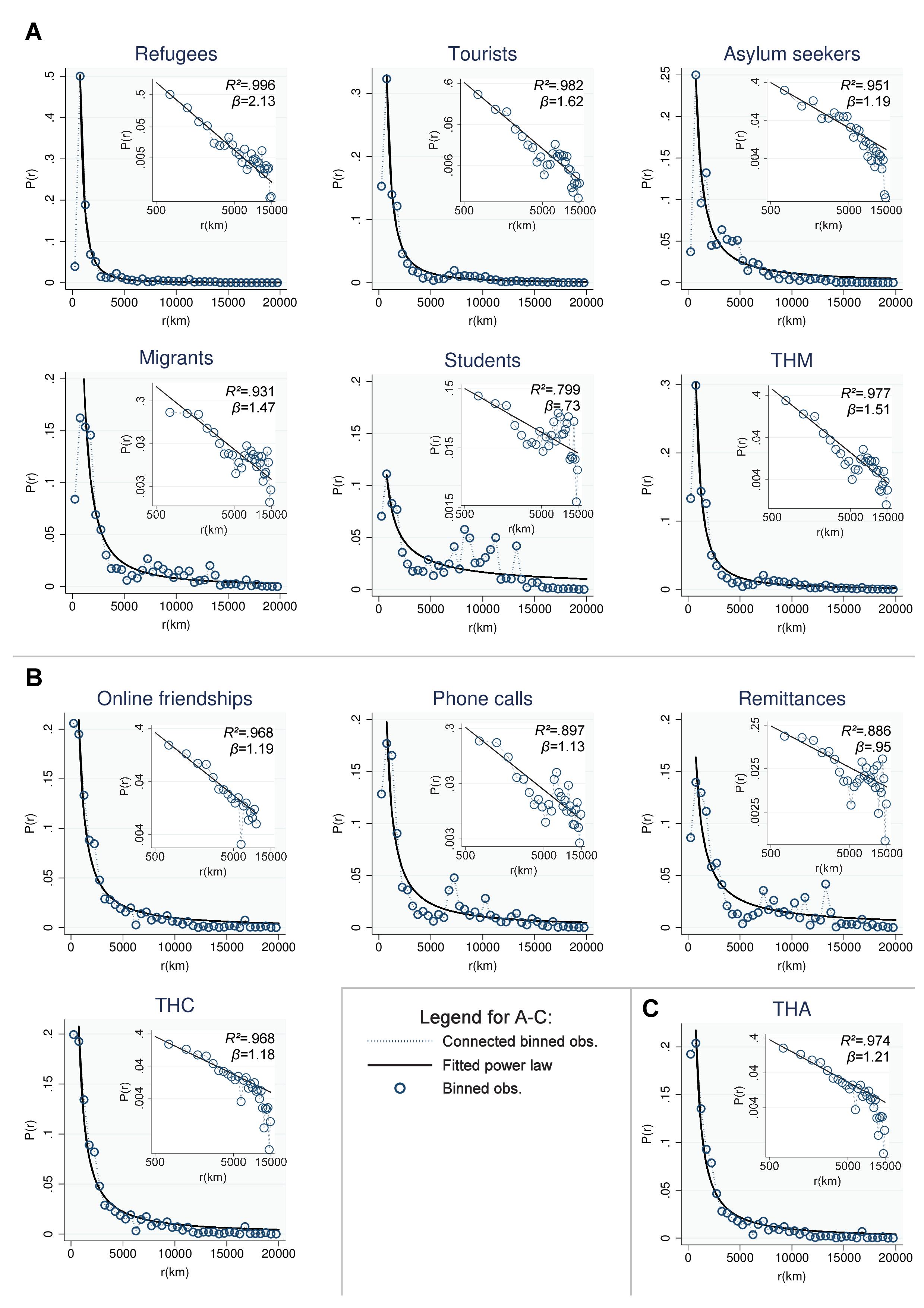}
\end{figure}

\subsection{Methods}
In order to determine the spatial structure of THA, we compare the empirically observed probability density functions of distances $r$ (in km) to the ideal pattern of a power-law distri-bution and describe to which extent the two are similar. To do so, we use the \textit{curvefit} module in Stata, which provides a goodness-of-fit measure ($R^2$) and other relevant parameters like the scaling exponent $\beta$. 

While the potential step lengths and directions are continuous in real physical space, here they are limited to a specific set based on the grid of the world's nation-states. In order to solve the issue that not every conceivable distance actually exists in this real-world set of country dyads (leading to spurious gaps) while some country pairs coincidentally feature the very same distance (resulting in spurious spikes), individual observations had to be binned first, i.e. step lengths that lie close to each other needed to be aggregated \citep[cf.][]{buchanan2008mathematical}. After trial computations with lower and higher step lengths, we decided to use a bin width of 500km as a reasonable compromise between inflating the variance and overly flattening the distribution. The binning results in a reduction of data points, i.e. the 38,220 original observations in each matrix are pooled into 39 meta-observations. In fitting the power law, we focus on the tail of the binned distribution (i.e. the part to the right of the global maximum, which lies at 500km for most activity types). This is necessary because activity at short distances (i.e. left of the global maximum) is artificially restricted as we focus on \textit{transnational} activity alone and cannot capture the huge amount of activity occurring within countries.

As a robustness check, we partially repeated the analysis with Kernel-weighted local poly-nomial smoothing using an Epanechnikov-kernel function and the ROT-bandwidth estimator, which minimizes the conditional weighted mean integrated squared error. This procedure, in which the number of observations remains at 38,220, leads to similarly high power-law fits (results available upon request). Due to space restrictions, we only present the findings from the more parsimonious binning method.
 \\

\section{Results}
We first examine the spatial structure of THA today, then turn to the analysis of longitudinal trends, and finally compare our findings to theoretical scenarios and results from other studies.

\subsection{The spatial structure of transnational human activity today}
Fig.~\ref{fig:one} shows the probability density distributions of all types of THA under study at the latest available point in time. The dots represent binned empirical observations, whereas the solid lines depict power-law curves that are fitted to the tails of the empirical distributions. The goodness-of-fit is indicated by the $R^2$ in the upper right corner (0=no fit, 1=perfect fit), by which the subgraphs are sorted. The scaling exponent $\beta$, placed below the $R^2$, indicates the steepness of the power-law curve. To additionally illustrate the power-law fit, the insets show the distributions and curves on logarithmic axes (on which the power law takes the form of a straight line along which the observations should cluster), for displacement lengths of $r\leq15,000$km. 

Concerning THM (Fig.~\ref{fig:one}A), the power-law fit is highest for refuge-seeking ($R^2$=.996), and just slightly lower for tourism ($R^2$=.982), asylum-seeking ($R^2$=.951), and migration ($R^2$=.931). Only for student exchange does the power law not fit the empirical distribution well ($R^2$=.799). The spatial structure of student exchange differs from that of the other THA types in that there is a second peak at middle-range distances (approx. 7,000-14,000km). A closer look at the data reveals that this peak results mainly from large flows between three country pairs (China--US, India--US, and Australia--UK), which fits our assumption about the role of the global university system's heavy stratification and the related lack of intervening opportunities. The scaling exponent $\beta$ is highest for refuge-seeking ($\beta$=2.13), medium-high for tourism ($\beta$=1.62), medium for migration ($\beta$=1.47), and small for asylum-seeking ($\beta$=1.19) and student exchange ($\beta$=0.73). Apart from the unexpected low $\beta$ for asylum-seekers\footnote{The unexpected values for asylum-seeking should not be overinterpreted as the number of asylum-seekers is relatively low, making the structure susceptible to rather meaningless fluctuations.}, the theoretically expected order holds, indicating that type-specific goal-attainment opportunities and available resources help predict spatial structures of human mobility across nation-state borders. THM as a whole also clearly follows a power law ($R^2$=.977, $\beta$=1.51).

\begin{sidewaysfigure}[p]
  \caption{Trends over time}
  \label{fig:two}
  \centering
    \includegraphics[width=1\textwidth]{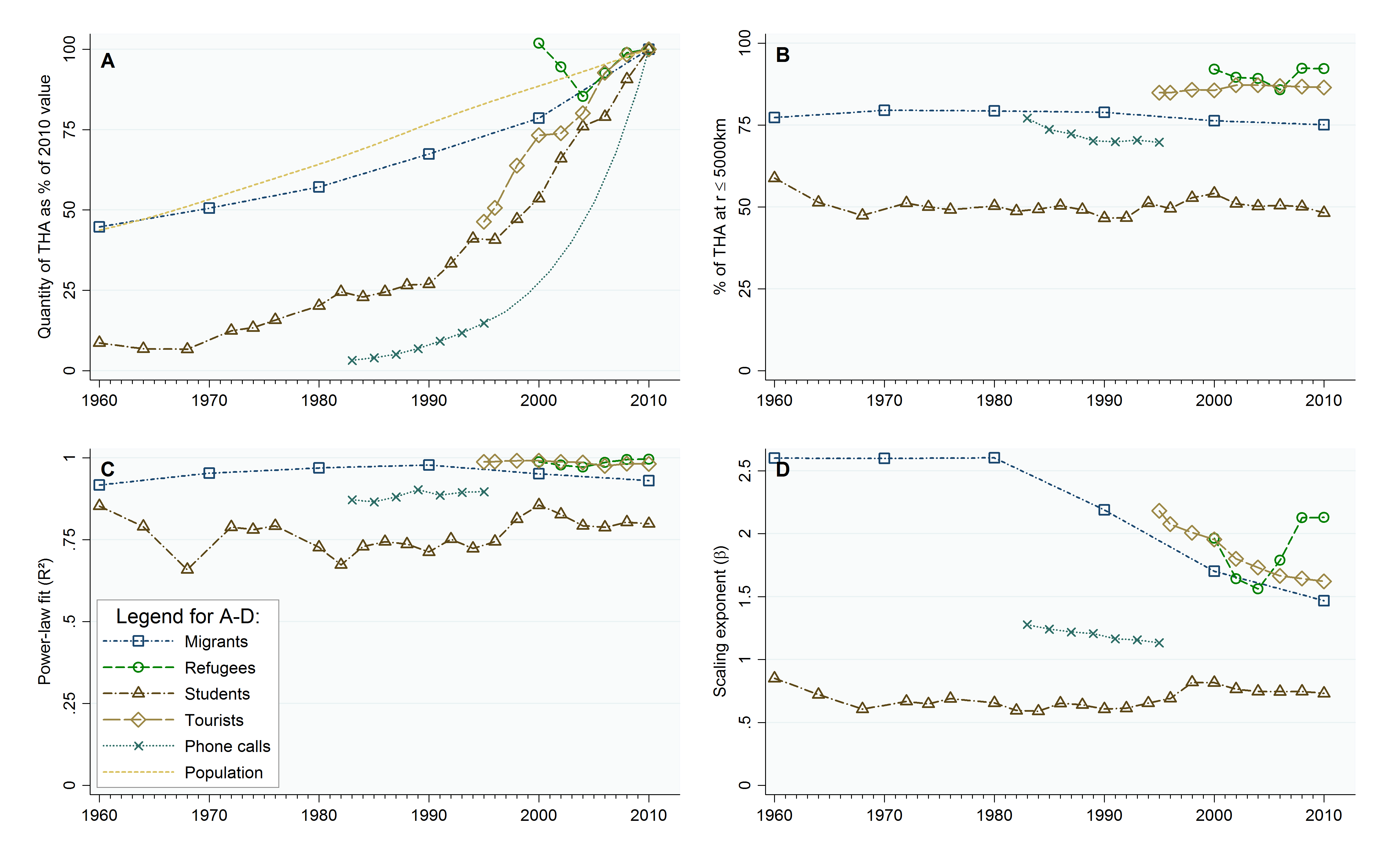}
\end{sidewaysfigure}

Regarding THC (Fig.~\ref{fig:one}B), the power-law fit and scaling exponent are highest for online friendships ($R^2$=.968, $\beta$=1.19), lower for phone calls ($R^2$=.897, $\beta$=1.13), and smallest for re-mittances ($R^2$=.886, $\beta$=0.95). The smaller scaling exponent for analogue phone calls in 1995 compared to digital Facebook friendships in the late 2000s is clearly at odds with the popular argument that lower communication costs lead to a ``death of distance'', or at least to a \textit{flatter} probability density distribution --- surprisingly, our data suggests the opposite. 

With respect to THA as a whole (Fig.~\ref{fig:one}C), we again find an excellent power-law fit ($R^2$=.974, $\beta$=1.21), indicating that most human activity beyond nation-state borders today only spans relatively short distances and genuinely global mobility and communication is still extremely rare. Contrary to popular notions about the detachment of human activity from spatial constraints in the age of globalization in the geography-is-dead debate, the forces of physical space seem to be intact. But is THA today, compared to the past, at least \textit{somewhat} more globalized? To answer this question, we now turn to analyzing trends over time.
 \\

\subsection{Developments over time}

To investigate longitudinal trends, we look at four different indicators: the overall amount of THA (Fig.~\ref{fig:two}A), the percentage of THA that takes place at relatively short distances (Fig.~\ref{fig:two}B), the power-law fit (Fig.~\ref{fig:two}C), as well as the scaling exponent $\beta$ (Fig.~\ref{fig:two}D).

Fig.~\ref{fig:two}A illustrates that almost all types of THA have seen massive growth over the years. The number of transnationally mobile students (depicted by triangles) grew exponentially from 255,000 in 1960 to 2.9 million in 2010, an 11.5-fold increase. Transnational phone-call minutes (depicted by crosses\footnote{For phone calls, post-1995 values are extrapolated by fitting an exponential trend line ($R^2$=.998) of the form $y=1E+10e^{0.13x}$, where $x$ is 1 in 1983, 2 in 1984, etc. This assumption of exponential growth, which leads to an estimate of 380.9 billion minutes for 2010, is conservative given that \citet[p. 2]{telegeography2014report} estimates the international call volume to exceed 400 billion minutes in 2010.}) increased from 12.4 billion in 1983 to 56.6 billion in 1995, rising 4.6-fold in little more than a decade. Similarly, the number of transnational tourists (depicted by diamonds) rapidly grew from 457.6 million in 1995 to 987.3 million in 2010, an increase of 115.8 percent. The number of migrants (depicted by squares) was estimated to be 91.2 million in 1960 and more than doubled (plus 123.6 percent) to 204.0 million in 2010. Only the number of refugees (depicted by circles) does not follow a clear direction over time. All in all, the amount of THA has increased dramatically\footnote{One may want to consider that the world population also increased over time: in the 196 countries under study, it grew from 3.0 billion in 1960 to 6.9 billion in 2010 (short-dashed line in Fig.~\ref{fig:two}A). Student mobility, tourism and phone calls grew at a faster rate, but, contrary to what catchphrases like ``age of migration'' \citep{castles2005age} suggest, the relative strength of migrants as a share of the world population decreased during the 1960s, '70s, '80s, and '90s, and only in 2010 reached the level it had in 1960.} over the last decades. 

By contrast, Fig.~\ref{fig:two}B-C show a remarkable degree of over-time stability when it comes to the spatial structure of THA. Fig.~\ref{fig:two}B illustrates the percentage of THA that takes place at displacement lengths of $r\leq5,000$km, i.e. at distances shorter than about \sfrac{1}{4} of the largest possible distance between two countries. For comparison: the driving distance between Seattle and Miami is 5,400km. We are thus talking about relatively short distances at the planet scale. The graph reveals two things: (a) the percentage of mobility and communication that takes place below this threshold varies between types of THA, and (b) it remains strongly attached to these type-specific levels over time. For students, the percentage of stays abroad at distances of $r\leq5,000$km remained at about 50.3 percent (standard deviation $\sigma=2.6$ percent) over a period of fifty years. Similarly, in every decade from 1960 to 2010, about 77.7 percent ($\sigma=1.8$ percent) of migrants moved these relatively short distances. At nine measured time points between 1995 and 2010, the fraction of tourists that went such short distances varied closely around 86.2 percent ($\sigma=0.9$ percent). For refugees, the share was about 90.2 percent ($\sigma=2.5$ percent). Only phone calls (mean: 71.9 percent, $\sigma=2.7$ percent) appear to witness a slow but constant drop -- yet the development of refuge-seeking during the 2000s shows how quickly such a trend can reverse. Overall, the consistently low standard deviations highlight the robustness of the spatial structure of THA over time. Similarly, the power-law fit (Fig.~\ref{fig:two}C) remains remarkably stable. For phone calls, it remains solid at about .886 ($\sigma=.014$). In the case of migration, it varies steadily around .950 ($\sigma=.023$). Tourism and refuge-seeking remain at the ceiling, with fits of .986 ($\sigma=.005$) and .986 ($\sigma=.009$) respectively. For student exchange there are some ups and downs, but all in all the power-law fit oscillates around .768 ($\sigma=.052$). Thus, the power-law fit is another indicator of the startling long-term stability of the spatial structure of THA. 

\begin{figure} [t]
  \caption{Log-log plots of THA over time}
  \label{fig:three}
%\floatfoot{\emph{Note}: Both axes are logarithmic.}
  \centering
    \includegraphics[width=1\textwidth]{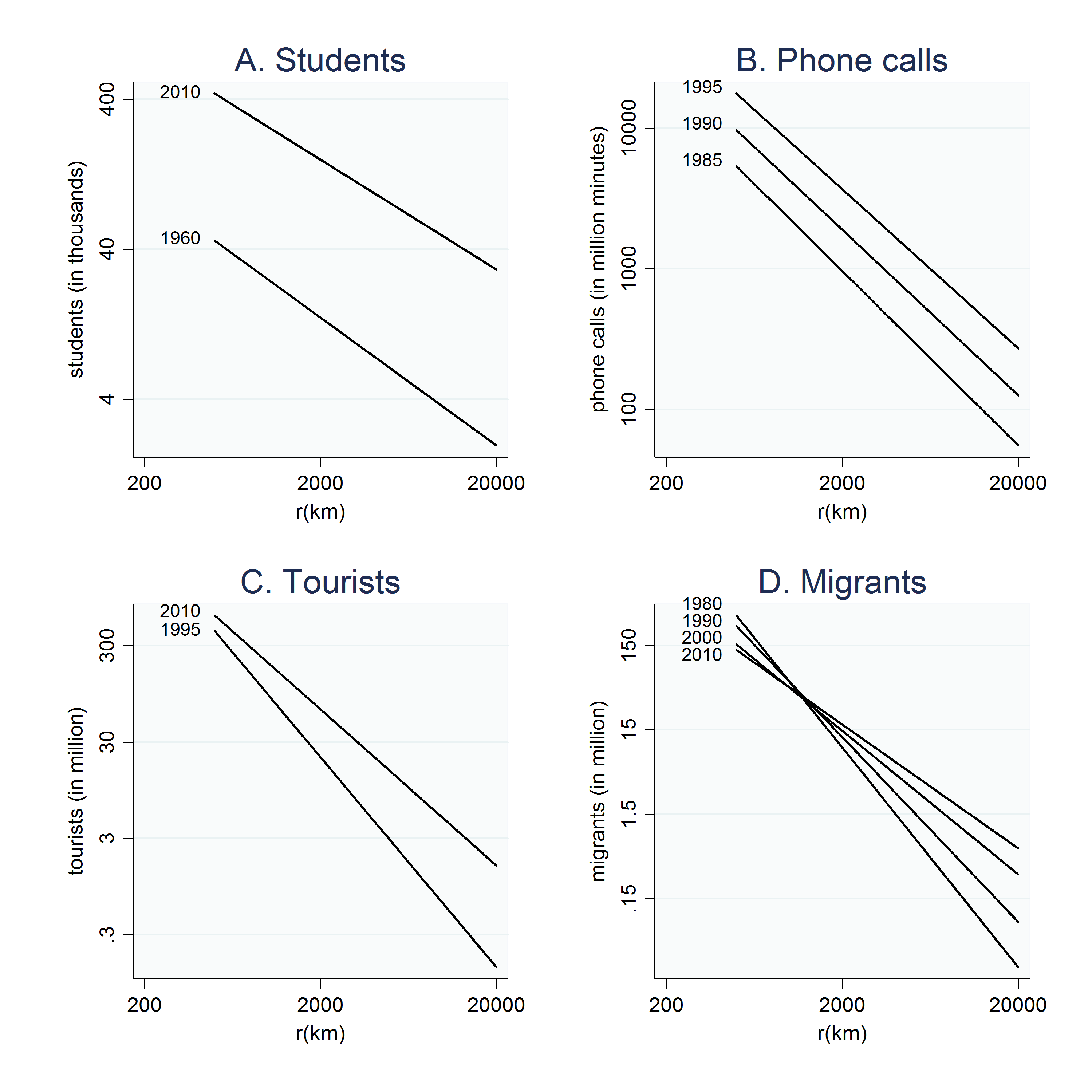}
\end{figure}

Per contra, Fig.~\ref{fig:two}D shows a mixed picture when it comes to the development of the scaling exponent $\beta$, i.e. the ``flatness of the world''. For tourists and migrants (which together represent 98.9 percent of all THM) a steady drop in the size of $\beta$ can be observed. For migration, this decrease is first visible in the 1980s, after two decades of absolute stability. With respect to phone calls, we see only a slight gradual decrease over time. Concerning refugees, $\beta$ first drops between 2000 and 2004, but then increases again until 2010. For students on the other hand, the scaling exponent remains more or less stable at a low level. This finding of partial stability and partial flattening can be examined more closely in log-log plots (Fig.~\ref{fig:three}). 

The graphs in Fig.~\ref{fig:three} show the best-fitting power-law lines (which form straight lines here, as both axes are logarithmic) for the spatial structures of the four types of THA with a clear $\beta$-value trend, at various points in time. To visualize both absolute growth and relative shifts, the y-axes now show the actual amount of mobility and communication taking place. Remarkable differences become apparent: for students and phone calls (Fig.~\ref{fig:three}A-B), the fitted power-law curves form parallel lines that just move upwards on the y-axes as the number of students and phone-call minutes increases over time. In other words, student exchange and phone calls grow at more or less the same rate at all distances over time, thus retaining their spatial structures' shapes. For tourists and migrants (Fig.~\ref{fig:three}C-D) however, the picture looks different. Here, the lines still move upwards, but also become flatter over the years. This difference shows that no universal trends hold for all types of THA, underlining the fruitfulness of our comparative approach.

At first sight, there appears to be a contradiction between the over-time stability observed in Fig.~\ref{fig:two}B and the shifts unveiled in Fig.~\ref{fig:three}C-D. Are at least some types of THA ``globalizing'' over time or not? The seeming paradox can be resolved by reminding oneself that the straight lines in Fig.~\ref{fig:three} are only imaginary --- in reality (i.e. on regular axes) they form heavily bended curves. Shifts that are meaningful in size only occur within the short-distance range (i.e. at displacement lengths of  $r\leq5,000$km), leaving the share of long-distance THA practically unaffected. Therefore, these changes should rather be taken as signs of regionalization (i.e. gradual extensions of the spatial reach of THA at a regional scale) than as evidence for globalization. At the planet-wide scale, stability prevails --- even for tourism and migration (as visible in Fig.~\ref{fig:two}B).
 \\

\subsection{Relation of our results to theoretical scenarios and other studies}
\begin{figure} [t]
  \caption{The relation of empirical THA to theoretical scenarios}
  \label{fig:four}
  \centering
    \includegraphics[width=0.5\textwidth]{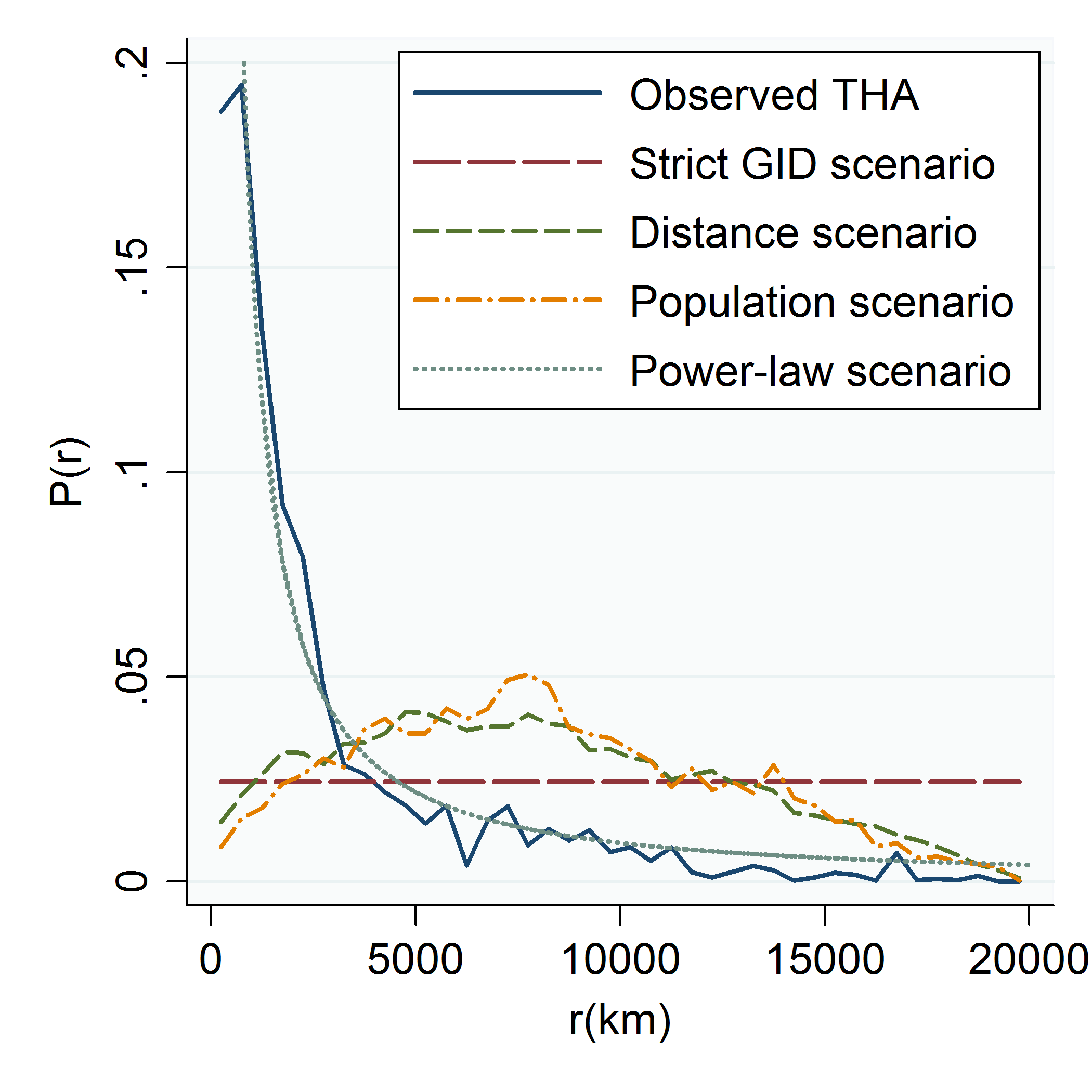}
\end{figure}

So far, we have mainly adhered to the L\'{e}vy-flight debate's approach in our analysis and only made allusions to the geography-is-dead debate en passant in the longitudinal analysis by assuming a ``flatten\textit{ing}'' world. In this last analytical step, we examine what would happen if the geography-is-dead debate's statements were understood in a stricter sense and distance played no role \textit{at all} anymore. To do so, Fig.~\ref{fig:four} compares the empirically observed spatial structure of THA (our overall index, depicted by the solid line) to various hypothetical scenarios. 

The first one, the \textit{strict geography-is-dead scenario}, shows what the distribution would look like if the probability of activity to occur would literally be the same at all distances (a ``flat world'' in Friedman's terms). The second one, the \textit{distance scenario}, refines this assumption by taking the actual geographic distribution of country dyads in the world into account. As short and long distances are empirically less common than middle-range distances, the relation would be reverse-u-shaped if the amount of activity was the same between each of the 38,220 existing country pairs (reminiscent of Marx and Engels's ``universal interdependence of nations''). The third one, the \textit{population scenario}, is a further refinement that takes into consideration that countries differ by population. It shows what would happen if each individual in each country had the same chance of going to any other country independent of its geographic location (akin to Toffler's new nomads who ``leave the physical structure behind''). 

It immediately becomes apparent that none of the three scenarios comes anywhere near the actual data. In reality, far more THA occurs at short distances and far less at longer distances than the geography-is-dead hypothesis in its strict sense would suggest. The dotted power-law line, which we include here again for comparison, shows that the L\'{e}vy-flight is much better at realistically representing the spatial structure of THA than any geography-is-dead scenario. Again, we find that distance is not dead at all.

Finally, we are also able to compare our findings to those of other studies at lower geographic scales and for different species. Concerning local human mobility, \citet{gonzalez2008understanding} report a scaling exponent of $\beta$=1.75, while \citet{brockmann2006scaling} specify $\beta$=1.59 for national human mobility in the US. Our value for transnational human mobility is only slightly lower with $\beta$=1.51 and practically identical to the $\beta$=1.50 reported by \citet{noulas2012tale} for the global structure of LSS log-ins, which may be taken as an indicator of robustness. For a set of marine predators, \citet{sims2008scaling} found $\beta$=2.12, whereas for spider monkeys, \citet{ramos2004levy} name $\beta$=2.18. This consistent ranking of the scaling exponents could indicate that while mobility follows a L\'{e}vy-flight pattern at all geographical scales and across many species, the scaling exponent is smaller (i.e. longer distances are relatively more common) in humans than in non-human species and at higher geographic scales than at lower geographic scales. With due caution not to over-interpret these numbers (which are known to be susceptible to fluctuation), we leave this impression to further scrutiny by future research.  
 \\\\

\section{Summary and conclusion}
Starting from contradicting propositions of largely disconnected debates in the natural and social sciences, this article comparatively analyzed the worldwide spatial structure of transnational human activity (THA) across eight different types of mobility and communication and in its development over time. Four findings should be highlighted:
\begin{enumerate} \itemsep1pt \parskip0pt
\parsep0pt
  \item Contrary to popular social scientific accounts, geography is not ``dead''. The large majority of THA still occurs at short distances and truly global mobility and communication continues to be scarce. The spatial structure of THA is similar to that of animal displacements and local-scale human motion in that it can be approximated by L\'{e}vy flights with heavy tails that obey power laws. 
  \item Scaling exponent and power-law fit differ by type of mobility, being highest in refuge-seeking and tourism and lowest in student exchange. This pattern suggests that the availability of opportunities for attaining type-specific goals as well as the resource stock disposable on average to the individuals engaging in a specific mobility type play a role in determining the precise shape of the spatial structure.
  \item For communication, we observe that distance is stronger in determining online friendships in the late 2000s than analogous phone calls more than a decade earlier. This finding is completely at odds with Cairncross's popular argument that declining costs of communication lead to a ``death of distance''.
  \item Despite dramatic increases in the absolute amount of transnational mobility and communication over the years, the L\'{e}vy-flight pattern remains intact and remarkably stable over time. Longitudinal change occurs only for some types of THA and predominantly at short distances, indicating shifts at the regional level rather than globalization. 
\end{enumerate}
Overall, our findings suggest that geography still shapes the patterns of planet-scale human activity --- be it physical mobility or indirect communication. Where humans interact across nation-state borders, they are very likely to do so with neighboring countries and within the world regions they live in. The discrepancy between our findings and popular accounts of globalization --- from McLuhan's ``global village'' to Friedman's ``flat world'' --- may indicate that we as humans tend to overestimate our capability to transcend nature. Despite the profound technological, socio-economic and infrastructural revolutions that took place during the last half century --- from the end of the bipolar world system after the fall of the iron curtain to the rise of mass tourism and the dawn of the internet age --- the overall patterns of human mobility remain largely unchanged, bound to the same natural laws as the motion of all kinds of species, from plankton to penguin. 

With this study, we aimed at filling a gap in the L\'{e}vy-flight debate by conducting an encompassing analysis of the spatial structure of planet-scale human activity, and at complementing the geography-is-dead debate by subjecting some of its propositions to a rigorous empirical test. Now we can say that our findings fit the L\'{e}vy-flight model well, whereas the death-of-distance arguments turned out not to hold under closer scrutiny for individual human mobility and communication. This result underlines the need for sociologists to engage in research trends beyond the boundaries of their own discipline. We deem further exchange concerning this topic across natural- and social-scientific camps highly desirable. 

There are many paths future empirical research could take: for one thing, it should be gainful to abandon the global perspective to comparatively examine the spatial structure by world region or separately for the world system's center and periphery. Another productive next step would be to compare the over-time stability of THA with the longitudinal systemic stability found in other domains, e.g. in the tripartite global wealth structure \citep{babones2005country}. Furthermore, future research should aim at going beyond the L\'{e}vy-flight debate's simplistic assumption of a random walker. As the directions of human activity are actually not random \citep{rhee2011levy, song2010modelling}, it needs to be scrutinized which factors (economic, political, legal, cultural, etc.) influence where humans go. Our analysis of the role of opportunities for type-specific goal-attainment and the average amount of available resources may be seen as a sketchy first step into this direction that needs to be further expanded.\footnote{In fact, this paper is part of a larger project in which this question is tackled.} Finally, it would be desirable to link our insights with other fields of research that are heavily intertwined with THA, from global food supplies and transportation to electricity transmission and the spread of epidemics. We share the hope of \citet{hui2010planet} that new insights into how we as humans move and communicate worldwide may eventually help tackle such pressing global problems.
 \\\\\\\\\\\\

\begingroup
\parindent 0pt
\def\enotesize{\normalsize}
\pdfbookmark{Endnotes}{Endnotes}
\theendnotes
\endgroup

%----------------------------------------------------------------------------------------
%	REFERENCE LIST
%----------------------------------------------------------------------------------------

%\bibliographystyle{abbrvnat}  % needs package natbib
\bibliographystyle{apsr}  % needs package natbib
\bibliography{\jobname}       % uses \jobname.bib, according to \jobname.tex

%----------------------------------------------------------------------------------------

%\end{multicols}

\end{document}